\documentclass[10pt]{article}

\usepackage[english]{babel}

\usepackage[a4paper,top=2cm,bottom=2cm,left=3cm,right=3cm,marginparwidth=1.75cm]{geometry}

\usepackage[backend=biber,style=nature]{biblatex}
\usepackage{amsmath}
\usepackage{amssymb}
\usepackage{graphicx}
\usepackage[colorlinks=true, allcolors=blue]{hyperref}
\usepackage[auth-sc, affil-it]{authblk}
\usepackage{caption}
\usepackage{subcaption}
\usepackage{csquotes}

\usepackage{booktabs}
\usepackage{colortbl}
\usepackage{xcolor}
\usepackage{xfrac}
\newcommand{\ra}[1]{\renewcommand{\arraystretch}{#1}}

\usepackage{parskip}
\providecommand{\keywords}[1]
{
  \small	
  \textbf{\textit{Keywords---}} #1
}

\title{Generating artificial displacement data of cracked specimen using physics-guided adversarial networks}
\author[1,*]{David Melching}
\author[1]{Erik Schultheis}
\author[1]{Eric Breitbarth}
\affil[1]{German Aerospace Center (DLR), Institute of Materials Research, Linder Hoehe, 51147 Cologne, Germany.}
\affil[*]{Corresponding author: David.Melching@dlr.de}

\date{December 6, 2023}

\begin{document}
\maketitle

\begin{abstract}
Digital image correlation (DIC) has become a valuable tool to monitor and evaluate mechanical experiments of cracked specimen, but the automatic detection of cracks is often difficult due to inherent noise and artefacts. 
Machine learning models have been extremely successful in detecting crack paths and crack tips using DIC-measured, interpolated full-field displacements as input to a convolution-based segmentation model. Still, big data is needed to train such models. However, scientific data is often scarce as experiments are expensive and time-consuming. 
In this work, we present a method to directly generate large amounts of artificial displacement data of cracked specimen resembling real interpolated DIC displacements. The approach is based on generative adversarial networks (GANs). During training, the discriminator receives physical domain knowledge in the form of the derived \textsc{von Mises} equivalent strain. We show that this physics-guided approach leads to improved results in terms of visual quality of samples, sliced Wasserstein distance, and geometry score when compared to a classical unguided GAN approach.
\end{abstract}


\section{Introduction} \label{sec:intro}

Fatigue crack growth (FCG) experiments are of significant importance to determine the lifetime and damage tolerance of critical engineering structures and components that are subjected to non-constant loads \cite{Tavares2017}. In recent years, digital image correlation (DIC) has been used to accompany and evaluate such mechanical experiments \cite{sutton2015recent}. The DIC data serves as the basis for subsequent mechanical evaluation of fracture mechanical quantities like the stress intensity factors  \cite{Roux2009DIC} and $J$-integral \cite{Becker2012DIC}. For this evaluation, the spatial location of the crack and especially the exact crack tip position is crucial. However, DIC data is subject to inherent noise and artefacts due to influences such as pattern quality, sensor noise, air movement, etc. \cite{Zhao2019} which makes this information difficult to obtain. To improve upon these issues, extensive work was done optimizing the pattern quality \cite{Chen2018} as well as the DIC algorithm in order to obtain reliable measurements in case of inferior patterns \cite{su_recursive-iterative_2020} or under special external conditions \cite{su_refractive_2021}. In the context of fracture mechanics, convolutional neural networks have been successfully applied to solve the crack detection problem fully automatically \cite{Strohmann2021, Melching2022}. These networks can deal with DIC noise and take the interpolated DIC-measured displacement fields as input to predict the crack paths and tips. However, for these powerful data-driven models to work reliably, they need a diverse set of training data. At the same time, experimental training data is scarce, since experiments are expensive, and manual labelling is extremely tedious and time-consuming. To address this issue, \textsc{Strohmann} et al. \cite{Strohmann2021} added artificial training data in the form of finite element (FE) simulations. Nevertheless, simulations are idealized and lack the characteristic DIC noise. 

Classically, synthetic DIC data can be generated by first creating an artificial speckle pattern on a digital image of the desired specimen. Then, an FE simulation is used to virtually deform the image. Finally, the deformed (speckle) image can be evaluated using a DIC algorithm (see, e.g., \cite{Lava2009} and more recently \cite{Rohe2022}). This synthetic DIC data can for instance be used to assess systematic errors arising from different DIC techniques \cite{Lava2009} or to understand calibration uncertainty \cite{Balcaen2017}. In contrast to these approaches, our goal is to bypass this multi-step procedure altogether and directly generate large amounts of artificial interpolated DIC-like training data in a fast, simple, data-centric manner using machine-learned generative models. For instance, this interpolated DIC data can be used to improve the training of automatic crack detection models (cf. \cite{Strohmann2021, Melching2022}) by delivering an advanced data augmentation method.

In the field of data-driven modeling, generative adversarial networks (GANs) have proven to be powerful data generators. GANs are a generative, unsupervised approach to machine learning based on deep neural networks trained using an adversarial process. Deep convolutional GANs (DC-GANs) produced state-of-the-art results in many generative and translative computer vision tasks such as image generation and style transfer \cite{Karras_2018_progressive, Karras_2020_CVPR, Gatys_NeuralStyleTransfer}.
However, training GANs typically requires large amounts of data, which are often not available in the scientific domain. For example, it is not possible to mechanically test an entire aircraft to generate training data. But without sufficient data, deep learning models are often unreliable and poorly generalize to domains not covered by the training data. To overcome this problem, efforts have been made to integrate fundamental physical laws and domain knowledge into machine learning models. \textsc{Karpatne} et al. \cite{Karpatne17TGDS} describe theory-guided data science as an emerging paradigm aiming to use scientific knowledge to improve the effectiveness of data science models. They presented several approaches for integrating domain knowledge in data science. 
\textsc{Daw} et al. \cite{Daw17PGNN} proposed a physics-guided neural network (PGNN) by adding a physics-based term to the loss function of a neural network to encourage physically consistent results.
\textsc{Karniadakis} et al. \cite{Karniadakis2021PINNs} coined the term physics-informed neural networks (PINNs) - a deep learning framework which enables the combination of data-driven and mathematical models described by partial differential equations. 
\textsc{Yang} et al. \cite{Yang2019PIGGAN} combined this approach with GANs by adding a physics-based loss to the generator and, recently, \textsc{Daw} et al. \cite{Daw21PIDGAN} introduced a physics-informed discriminator GAN, which physically supervises the discriminator instead.

In this work, we generate artificial displacement data of cracked specimen using GANs. Our framework is based on the classical DC-GAN architecture. We incorporate mechanical knowledge by using a physics-guided discriminator. In addition to the generated displacement data, this discriminator receives the equivalent strain according to \textsc{von Mises} \cite{vonMises1913} derived from the real or fake (generated) displacements as an additional input feature, see Section \ref{sec:method} and Figure \ref{fig:StrainGAN}. This mechanically motivated physical feature guides the adversarial training process and leads to physically consistent generated data. 
The generated data samples can be used to increase data variation of given training datasets consisting of interpolated DIC displacements. Although this synthetic data is not labelled, it has the potential to improve supervised machine learning tasks, e.g. by using unsupervised pre-training \cite{Erhan10UnsupervisedPretraining} or label propagation \cite{Iscen2019CVPR}.
In general, our method can be applied to generate interpolated DIC-like displacement fields of cracked specimen. In this paper, we focus on FCG experiments of a middle tension specimen manufactured from an aluminium-based alloy (see Section \ref{subsection:data}) and train several GANs.
To demonstrate the merits of the physics-guided method, we compare the results of the physics-guided GAN to a classical GAN approach in terms of visual quality of generated samples (see Section \ref{subsec:visual_eval}) and distance of fake data distributions to the real training data. The latter is quantified using the sliced Wasserstein distance (SWD) \cite{RabinWasserstein2012} and the geometry score (GS) \cite{GeometryScore} (see Section \ref{subsec:evaluation_methods}). 
We show that the physics-guided approach accelerates the training and leads to physically more consistent results.

\section{Methodology} \label{sec:method}

Generative Adversarial Networks (GANs) are generative machine learning models learned using an adversarial training process \cite{Goodfellow_GANs}. In this framework, two neural networks - the generator $G$ and the discriminator $D$ - contest against each other in a zero-sum game. 
Given a training dataset characterized by a distribution $p_{\text{data}}$, the generator aims to produce new data following $p_{\text{data}}$ while the task of the discriminator is to distinguish generated data samples from actual training data samples. 

Given a noise vector $z$ sampled from a prior, e.g. the standard normal distribution, the generator outputs data samples $G(z)$, called \textit{fake samples}, trying to follow the training data distribution $p_{\text{data}}$.
Given a real or fake sample, the discriminator is supposed to decide whether it is real or fake by predicting the probability of it belonging to the training dataset.

Both models $G$ and $D$ are trained simultaneously contesting against each other in a two-player zero-sum minimax game with the value function $V(G,D)$:\
\begin{equation}
    \min_G \max_D V(G,D) := \mathbb{E}_{x \sim p_{\text{data}}} \left[ \log(D(x)) \right] + \mathbb{E}_{z} \left[ \log(1-D(G(z))) \right]
\end{equation}
This means $D$ is trained to minimize the discriminator loss
\begin{equation} \label{eq:discriminator_loss}
    L_D = - \log(D(x)) - \log(1-D(G(z))),
\end{equation}
whereas $G$ is trained to minimize the generator loss
\begin{equation} \label{eq:generator_loss}
    L_G = \log(1-D(G(z))).
\end{equation}

As the discriminator gets better at identifying fake samples $G(z)$, the generator has to improve on generating samples which are more similar to the real training samples $x \sim p_{\text{data}}$. We refer to \cite{Goodfellow_GANs} for further details of the training algorithm.

\subsection{Digital image correlation}
Digital image correlation (DIC) is a contact-less, optical method to obtain full-field displacements and strains. It is widely applied in science and engineering to quantify deformation processes. In experimental mechanics, it is used to monitor and evaluate fatigue crack growth (FCG) experiments \cite{MOKHTARISHIRAZABAD201611} by determining fracture mechanical parameters like stress intensity factors (SIFs) \cite{Roux2009DIC} or the J-integral \cite{Becker2012DIC}. Essentially, DIC measurements are based on the comparison of a current image with a reference image using tracking and image registration techniques. The cross correlation method requires a random speckle pattern on the sample surface. Various external and internal influences such as illumination, air movement, vibrations, facet size and spacing, camera settings, sensor noise, pattern quality, etc. lead to inherent noise in the DIC data. Our goal is to generate artificial interpolated DIC-like displacement data using GANs. Since this data incorporates characteristic DIC noise, it can subsequently be used to improve the training of machine learning models such as crack detection \cite{Strohmann2021, Melching2022}.

\subsection{Training data} \label{subsection:data}
To create a dataset for the training of our GANs, we use planar displacement fields $u=(u_x,u_y)$ obtained during FCG experiments of the aluminium alloy AA2024-T3 using a commercial GOM Aramis 12M 3D-DIC system. Details on the general experimental setup can be found in \cite{Strohmann2021}. For the dataset, we use \textit{one} FCG experiment performed on a middle tension (MT) specimen (width $w=160$\,mm, thickness $t = 2$\,mm). While the specimen is clamped at the bottom, a load is applied from the top with a maximal force of $F_{\text{max}}= 15$\,kN (corresponding to $\sigma_{\text{max}}= 47$\,MPa) and ratio $R=0.3$ with 20 load cycles per second. Every 0.5\,mm of crack growth (measured by direct current potential drop), 3 images (at minimal load $F_{\text{min}} = R\cdot F_{\rm max}$, mean load $F_{\text{min}}+(F_{\text{max}}-F_{\text{min}})/2$ and maximum load $F_{\text{max}}$) were acquired. From the resulting DIC dataset, we take the planar displacements $u_x$ and $u_y$ of the specimen and linearly interpolate them from an area of $70 \times 70 \,\text{mm}^2$ of the right-hand side of the specimen on an equidistant $256 \times 256$ pixel grid. This procedure results in 838 data samples of shape $2 \times 256 \times 256$, where the first dimension stands for the $x$- and $y$- displacements. Each of the two channels is normalized to $[-1, 1]$ by the min-max-scaling and shift
\begin{equation}
    u_{\rm scaled} = 2 \cdot \frac{u-u_{\rm min}}{u_{\rm max}-u_{\rm min}} - 1
\end{equation}
such that the minimum and maximum are mapped to $-1$ and $1$, respectively.

\subsection{Physics-guided GAN}
We aim to generate artificial interpolated DIC displacement data using deep convolutional GANs. For this, we mainly follow the architectural guidelines from \cite{DCGAN2015}. However, in order to reduce checkerboard artefacts, we choose nearest-neighbor upsampling instead of transposed convolutions in the generator \cite{odena2016deconvolution}. 
We remark that GANs cannot be expected to generalize beyond the training data. The reason for this is that the generator learns to produce fake samples that approximate the distribution of the training data. To cover a different type of experiment or material, the training data set must be extended.

\begin{figure}[ht]
    \centering
    \includegraphics[width=\textwidth]{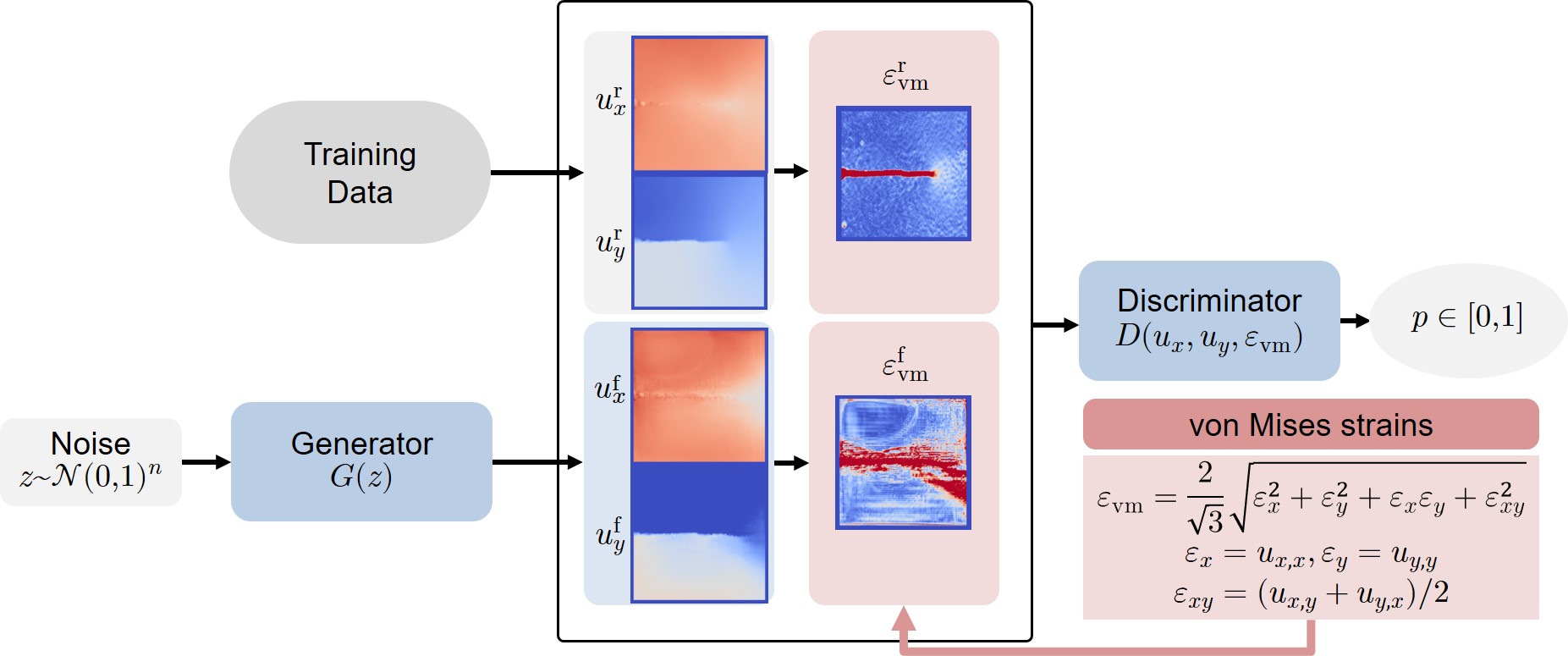}
    \caption{Physics-guided GAN framework: A deep convolutional generator $G$ creates fake interpolated DIC displacement data samples from noise $z$. These samples $(u_x^{\rm f}, u_y^{\rm f})$ are used to calculate the corresponding {\sc von Mises} equivalent strain $\varepsilon_{\rm vm}^{\rm f}$. All these three features are handed to the discriminator $D$, which has to decide whether samples are real or fake.}
    \label{fig:StrainGAN}
\end{figure}

\textit{Generator.} The input of the generator network is a $n$-dimensional vector $z$ randomly sampled from a standard normal distribution. For definiteness, we choose $n=5$ throughout all our training experiments. First, the random vector $z$ passes a fully-connected layer with $8 \cdot 8 \cdot 512 = 32768$ neurons, batch-normalization \cite{BatchNorm_paper}, and Rectified Linear Unit (ReLU) activation \cite{ReLUBoltzmannMachines_paper}. The output of this layer is then reshaped into 512 features of size $8 \times 8$. After that, these features are successively doubled in size using the base block (upsampling $\rightarrow$ batch normalization $\rightarrow$ ReLU $\rightarrow$ convolution) four times. The final block ends with a $\tanh$ activation instead of ReLU. Therefore, in accordance with the training data, the generator outputs \textit{fake} samples with pixel values between $-1$ and $1$.

\textit{Discriminator.} For the discriminator, we implemented the following two approaches:

\begin{enumerate}
    \item \textbf{Classical:} The discriminator gets real and fake pairs of interpolated $x$- and $y$- displacement fields $(u_x,u_y)$ and predicts a (pseudo-)probability of the sample being real. We refer to this approach as \textit{classical} GAN.
    \item \textbf{Physics-guided:} In addition to the interpolated displacement fields, the corresponding \textsc{von Mises} equivalent strain $\varepsilon_{\rm vm}$ is calculated based on the generated and real interpolated displacement fields and the discriminator gets the triple $(u_x, u_y, \varepsilon_{\rm vm})$ as input in order to decide whether it is fake or real. For small strains, the {\sc von Mises} equivalent strain is defined as the scalar quantity
    \begin{equation} \label{eq:vonMisesstrain3D}
        \varepsilon_{\rm vm} = \sqrt{\frac{2}{3} \varepsilon_{\rm dev} : \varepsilon_{\rm dev}}, \qquad \text{where } \varepsilon_{\rm dev} = \varepsilon - \frac{1}{3} {\rm tr}(\varepsilon)
    \end{equation}
    denotes the deviatoric part of the three-dimensional strain tensor
    \begin{equation}
        \varepsilon = \begin{pmatrix}
                            \varepsilon_{xx} & \varepsilon_{xy} & \varepsilon_{xz}\\
                            \varepsilon_{xy} & \varepsilon_{yy} & \varepsilon_{yz}\\
                            \varepsilon_{xz} & \varepsilon_{yz} & \varepsilon_{zz}\\
                    \end{pmatrix}
    \end{equation}
    In case of plane stress, $\varepsilon_{xz} = \varepsilon_{yz} = 0$ and $\varepsilon_{zz} = - \nu (\varepsilon_{xx} - \varepsilon_{yy})$. Assuming volume constancy with a Poisson ratio of $\nu = 1/2$, Formula \eqref{eq:vonMisesstrain3D} simplifies to
    \begin{equation} \label{eq:vonMisesstrain2D}
        \varepsilon_{\rm vm} = \frac{2}{\sqrt{3}} \sqrt{\varepsilon_{xx}^2 + \varepsilon_{yy}^2 + \varepsilon_{xy}^2 + \varepsilon_{xx}\varepsilon_{yy}},
    \end{equation}
    We use Formula \eqref{eq:vonMisesstrain2D} for the physics-guided discriminator. Therefore, we numerically approximate the strains using finite differences, e.g.
    \begin{equation}
        \varepsilon_{xy}(x,y) = \frac{\partial}{\partial y} u_x(x,y) =  \frac{u_x(x,y+h) - u_x(x,y)}{h} + \mathcal{O}(h).
    \end{equation}
    To guarantee differentiability, the square-root function is smoothed by using $\sqrt{\cdot + \delta}$ with $\delta \ll 1$. We refer to this approach as \textit{physics-guided} GAN (see Figure \ref{fig:StrainGAN}).
    The discriminator, which drives the training of the generator, has additional physical information, namely the equivalent strain, that the generator can only influence indirectly by producing physically consistent displacement fields. Certainly, other quantities, which can be derived from the displacement fields, such as strains $\varepsilon_{xx}$, $\varepsilon_{xy}$, or $\varepsilon_{yy}$ can be used to guide the discriminator. However, we decide on using the equivalent strains, since the crack path and crack tip field is well-visible in them.
\end{enumerate}

In both cases, we choose the same model architecture for the discriminator. The input of size $2 \times 256 \times 256$, or $3 \times 256 \times 256$ in case of the physics-guided discriminator, is successively downsampled to the size $1\times 32 \times 32$ using three blocks of strided convolutions, batch normalization, and LeakyReLU activation \cite{leakyrelu}, where $\text{LeakyReLU}(t) = \max (\alpha t, t)$ with $\alpha=0.2$. The extracted features are then flattened and pass the last fully-connected layer with one output neuron and sigmoid activation. The output is a number between 0 and 1 and is interpreted as the probability of the sample being real.

\subsection{Evaluation of GANs} \label{subsec:evaluation_methods}
In classical supervised learning, a model is trained by minimizing a specific loss (e.g. mean squared error), which quantitatively compares model predictions with the expected target. After training, models can be evaluated and compared by calculating the loss (and accuracy) for independent labeled test data. GAN generators, however, are trained in an adversarial fashion using a second model (the discriminator) to classify generated data as real or fake. Both models are trained simultaneously to maintain an equilibrium. Therefore, there is no natural objective measure to evaluate GANs, quantitatively. Instead they are evaluated by assessing the quality and variation of generated data. This is typically achieved by visual inspection of generated samples or by calculating the inception score (IS) \cite{IS} and Fr\'echet inception distance (FID) \cite{FID}. However, in case of interpolated DIC data, several domain experts would be needed to objectively grade the visual quality of generated samples. Moreover, quantitative metrics like IS or FID can only be employed for natural images since they use image classification networks like Inception \cite{Inception}, which are pre-trained on ImageNet \cite{ImageNet}. Therefore, in addition to a visual examination of generated samples in Section \ref{subsec:visual_eval}, we use metrics which are independent of the data type and do not use any pre-trained models. More precisely, we use the following two metrics:

\textit{Sliced Wasserstein distance.} 
In mathematics, the Wasserstein distance is a natural distance function between two distributions. Intuitively, it can be viewed as the minimal cost of transforming one of the distributions into the other. In case of image-like datasets $X=\{X_n\}_{n=1,\dots N}$ and $Y=\{Y_n\}_{n=1,\dots N}$ with same number of samples $N$ and image sizes $c \times h \times w$, where $c$ is the number of channels and $h$ and $w$ denote the height and width of images, respectively, the (quadratic) Wasserstein distance is given by,
\begin{equation} \label{eq:Wasserstein_distance}
W(X,Y)^2 = \min_{\pi} \sum_{i,j,k,n} |X_n(i,j,k)-Y_{\pi(n)}(i,j,k)|^2,
\end{equation}
where the minimum is taken over all permutations $\pi$ of the set $\{1,\dots N\}$ \cite{RabinWasserstein2012}. 
Due to the high dimensionality of images and the large number of samples, the exact computation of the Wasserstein distance is computationally infeasible. This is because the number of permutations scales exponentially with the number of samples $N$. Therefore, instead of \eqref{eq:Wasserstein_distance}, we use the sliced Wasserstein distance (SWD) introduced in \cite{RabinWasserstein2012} as an approximation, which is amendable for efficient numerical computation. The main idea of slicing is to map the high dimensional image data from $\mathbb{R}^{c\times h\times w}$ onto one-dimensional slices. On these slices, the Wasserstein distance can be calculated in loglinear time by using the ordered structure of one-dimensional Euclidean space. The sliced Wasserstein distance is defined as,
\begin{equation} \label{eq:SWD}
\tilde{W}(X,Y)^2 = \int_{\theta \in \Omega} \min_{\pi_{\theta}} \sum_{n=1}^N |\langle X_n-Y_{\pi_{\theta}(n)}, \theta \rangle|^2 d\theta,
\end{equation}
where $\Omega = \{\theta \in \mathbb{R}^{c \times h \times w} : \|\theta\|=1\}$ denotes the unit sphere. We refer to \cite{RabinWasserstein2012, Karras_2018_progressive} and Section \ref{subsec:SWD} below for further details.

\textit{Geometry score.}
Introduced by Khrulkov \& Oseledets \cite{GeometryScore}, the geometry score (GS) allows to quantify the performance of GANs trained on datasets of arbitrary nature. It measures the similarity between the real dataset $X_{\rm real}$ and a generated one $X_{\rm fake}$ by comparing topological properties of the underlying low-dimensional manifolds \cite{Goodfellow-et-al-2016}. The detailed quantitative characterization of the underlying manifold of a given dataset $X$ is usually very hard. 
The core idea of \cite{GeometryScore} is to choose random subsets $L \subset X$ called \textit{landmarks} and to build a family of \textit{simplicial complexes}, parametrized by a non-negative, time-like \textit{persistance parameter} $\alpha$. For small $\alpha$, the complexes consist of a disjoint union of points. Increasing $\alpha$ adds more and more simplicies finally leading to one single connected blob. 
For each value of $\alpha$, topological properties of the corresponding simplicial
complex, namely the number of \textit{one-dimensional holes} in terms of homology, $\beta_1(\alpha)$, are calculated (see, e.g., \cite{HatcherAlgebraicTopology}). From this, the authors propose to compute Relative Living Times (RLTs) for every number of holes that was observed \cite{GeometryScore}. For each non-negative number $i$, the RLT is the amount of the time when exactly $i$ holes were present relative to the overall time $\alpha_{\text{max}}$ after which everything is connected. More precisely,
\begin{equation} \label{eq:rlt}
    \textrm{RLT}(i,X,L) = \frac{\mu\left(\{\alpha \in [0,\alpha_{\text{max}}] : \beta_1(\alpha) = i\}\right)}{\alpha_{\text{max}}},
\end{equation}
where $\mu$ denotes the standard Lebesgue measure. Since the RLTs depend on the choice of landmarks $L$, we choose a collection of $n$ random sets of landmarks $L_j$ and define the Mean Relative Living Times (MRLTs) as
\begin{equation} \label{eq:mrlt}
    \textrm{MRLT}(i, X) = \frac{1}{n} \sum_{j=1}^n \textrm{RLT}(i,X,L_j).
\end{equation}
The MRLT is a discrete probability distribution over the non-negative integers. It can be interpreted as the probability of the manifold having exactly $i$ one-dimensional holes (on average). The $L^2$-distance between the MRLT distributions of $X_{\rm real}$ and $X_{\rm fake}$ defines a measure of topological similarity between the real dataset and the generated one, called geometry score (GS):
\begin{equation} \label{eq:GS}
    \textrm{GS}(X_{\rm fake}, X_{\rm real}) = \sum_{i=0}^{i_{\rm max}-1} \left| \textrm{MRLT}(i, X_{\rm fake}) - \textrm{MRLT}(i, X_{\rm real}) \right|^2,
\end{equation}
where $i_{\rm max}$ is an upper bound on the number of holes. We refer to \cite{GeometryScore} for further theoretical details and to Section \ref{subsec:GS} for the choice of hyperparameters and results in our case.

\section{Results and discussion} \label{sec:results}
In order to demonstrate the effectiveness of the method and to compare the classical with the physics-guided discriminator approach, we trained 10 randomly initialized classical and physics-guided GANs each for 100 epochs. Moreover, we trained two classical and physics-guided GANs each for 1000 epochs in order to compare both architectures after long training runs. The training setup is described in Section \ref{subsec:training_procedure} below. The trained models are evaluated qualitatively and quantitatively by using the following criteria:
\begin{itemize}
    \item Visual inspection of randomly generated samples (Section \ref{subsec:visual_eval})
    \item Sliced Wasserstein distances (Section \ref{subsec:SWD})
    \item Geometry scores (Section \ref{subsec:GS})
\end{itemize}
A summary of the results can be seen in Table \ref{table:summary_results}. In short, the physics-guided GAN approach leads to visually better results after 100 epochs and overall to measurably better results. For a detailed discussion, we refer to the sections below.

\begin{table}[ht]
    \centering
    \ra{1.4}
        \begin{tabular}{ccccc}\toprule
        Model & Epoch & Visual quality & GS $\times 10^3$ & SWD $\times 10^3$ \\
        \midrule
        \rowcolor{black!10} Classical & 100 & low & 183.49 & 151.70 \\
        Physics-guided & 100 & medium & \textbf{157.58} & \textbf{82.96} \\
        \rowcolor{black!10} Classical & 1000 & high* & 23.83 & 83.19 \\
        Physics-guided & 1000 & high* & \textbf{1.93} & \textbf{61.61} \\
        \bottomrule
        \end{tabular}
    \caption{Subjective visual quality, calculated geometry score (GS), and sliced Wasserstein distance (SWD) (lower is better) for different GAN model architectures and training lengths. (*) apart from \textit{garbage} samples (see Figure \ref{fig:samples_e1000})}
    \label{table:summary_results}
\end{table}

\subsection{Training procedure} \label{subsec:training_procedure}
Before training, the filters of the convolution layers in both generator and discriminator network are initialised randomly from a normal distribution with zero mean and a standard deviation $0.02$. In contrast, the weights of the batch normalization layers are initialised from mean $1$ and standard deviation of $0.02$, whereas the biases are initialised with zeros.

For training, we choose the Adam optimizer \cite{AdamArXiv} with a learning rate of $0.002$, momentum parameters of $\beta_1=0.5, \, \beta_2=0.999$, and a batch size of $8$. We noticed that occasionally models suffer from mode collapse during training. This means that the generator always outputs the same or visibly similar fake data samples and stops to learn. This problem is well-known and still part of active research. Popular strategies to overcome convergence issues of GANs regularize or perturb the discriminator \cite{Arjovsky17, Roth17} or by using a more sophisticated loss function \cite{WGANs}. In our case, if mode collapse happened, we restarted the training and discarded the collapsed model. All neural networks and training loops were implemented using PyTorch \cite{PyTorch}. The hardware for the training was an NVIDIA RTX8000 graphics card.

\subsection{Visual evaluation} \label{subsec:visual_eval}
We begin with a visual inspection of the generated data and compare real training data to representative samples generated by the classical GAN and the physics-guided GAN. We refer to fake samples generated by the classical or physics-guided GAN generators as classical or physics-guided GAN samples, respectively.

\begin{figure}
    \centering
    \includegraphics[width=\textwidth]{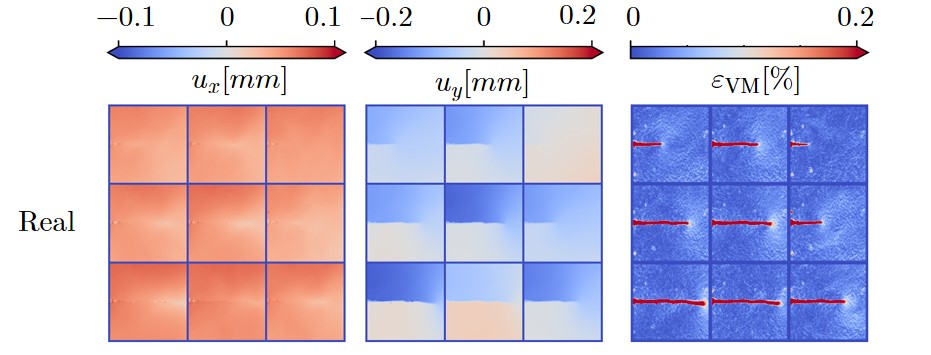}
    \caption{Random samples from the training dataset. Left: $x$-displacements. Middle: $y$-displacements. Right: \textsc{von Mises} equivalent strains. The corresponding grid elements belong to the same data point.}
    \label{fig:samples_real}
\end{figure}

Figure \ref{fig:samples_real} shows real interpolated DIC data samples obtained during FCG experiments as described in Section \ref{subsection:data}. The figure contains planar displacements and \textsc{von Mises} equivalent strains of 9 randomly selected data samples. Images belong together in the sense that the $x$-displacement of the first sample is located at the top left of the left column. The corresponding $y$-displacement is located at the same position in the middle column, and the corresponding calculated equivalent strain is located at the same position in the right column. Here, the crack path as well as the characteristic crack tip field is clearly visible. 

\begin{figure}
    \centering
    \includegraphics[width=\textwidth]{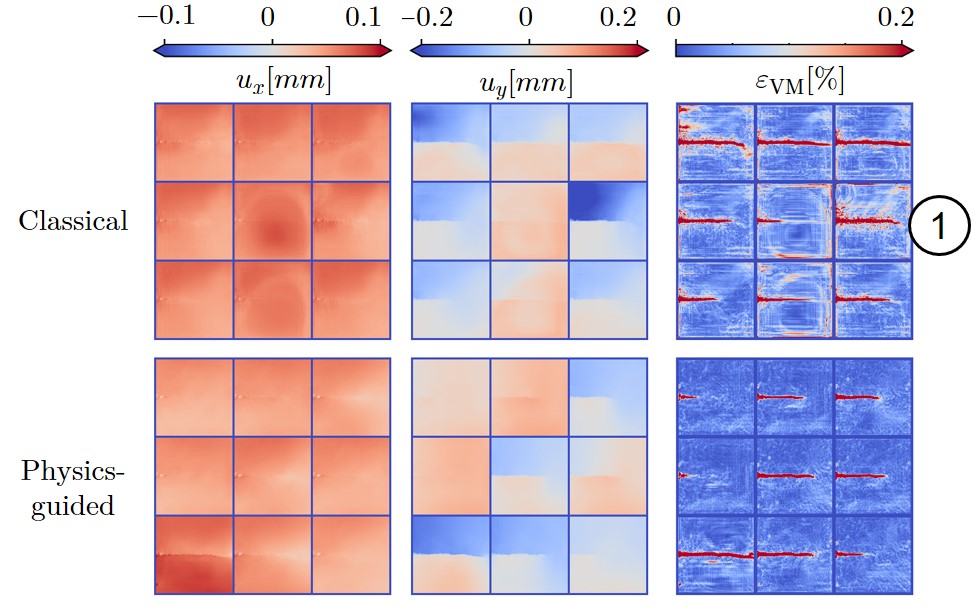}
    \caption{Visual comparison of randomly generated classical and physics-guided GAN samples after 100 epochs of training. Classical GAN samples show a larger noise level in the \textsc{von Mises} strains compared to phyiscs-guided GAN samples \textcircled{1}.}
    \label{fig:samples_e100}
\end{figure}

In Figure \ref{fig:samples_e100}, we see random fake samples after 100 epochs of GAN training. We can often identify the initial crack on the left edge and the crack path. Whereas most generated displacements are visually close to real displacements, significant differences are revealed in the \textsc{von Mises} strains, which are calculated afterwards. Especially classical GAN samples contain inconsistencies between $x-$ and $y-$ displacements and visual artefacts. This leads to large-scale vortexes and small-scale noise in the \textsc{von Mises} strains \textcircled{1}. Although far from being perfect, physics-guided GAN samples contain significantly less of these artefacts and inconsistencies and visually capture the inherent noise of the DIC system much better than classical GAN samples. Nevertheless, most fake samples are still visually distinguishable from real samples. In order to make sure the models are fully converged, we also performed some longer training runs.

\begin{figure}
    \centering
    \includegraphics[width=\textwidth]{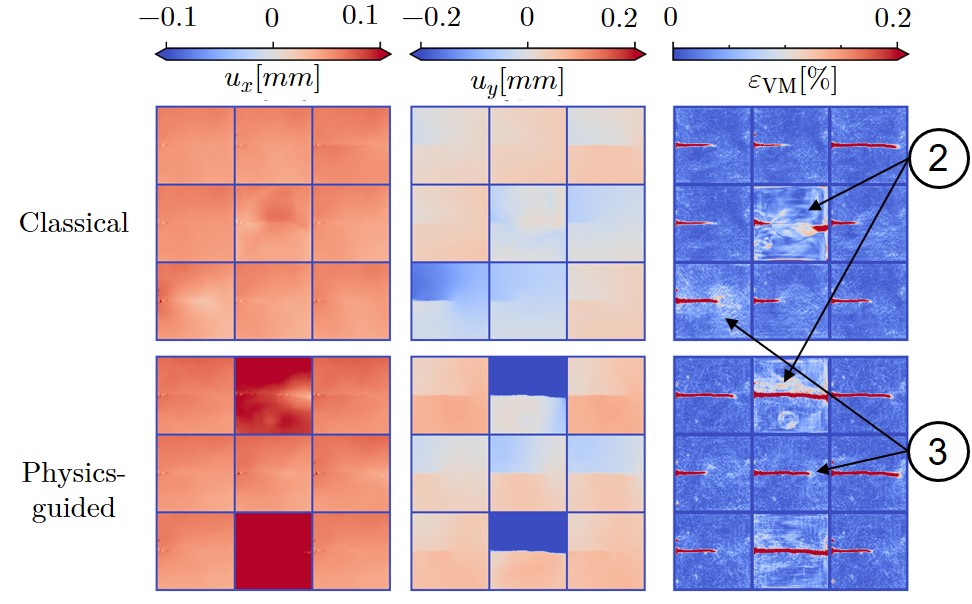}
    \caption{Visual comparison of randomly generated classical and physics-guided GAN samples after 1000 epochs of training. Both models seem to produce mostly good samples \textcircled{3} but also few garbage samples \textcircled{2}.}
    \label{fig:samples_e1000}
\end{figure}

Figure \ref{fig:samples_e1000} shows random fake samples after 1000 epochs. At this stage, the models are well converged and the visual difference between classical and physics-guided GAN samples has mainly disappeared. In general, the fake samples of both models show much better visual quality and less artefacts and inconsistencies compared to fake samples of generators trained for only 100 epochs \textcircled{3}. However, few samples suffer from severe inconsistencies and are qualitatively inferior \textcircled{2}. We refer to these failures as \textit{garbage} samples. Garbage samples may arise from difficulties in the training process of GANs and problems of non-convergence like mode collapse, which is an open research problem \cite{IS}. Although we do not observe mode collapse for the results shown here, the occurrence of garbage samples is a sign of local non-convergence, i.e. some noise inputs are mapped to garbage samples. Since the models are converged after 1000 epochs w.r.t. the difference in output between two epochs, the garbage samples seem to originate from intrinsic difficulties in the training process and are not related to the number of training epochs. 
Apart from these outliers, the vast majority of samples (of both models) are visually indistinguishable from real samples. Nevertheless, domain experts may notice that the characteristic crack tip field still seems unphysical in the fake samples especially when compared to real samples with long cracks.

\subsection{Sliced Wasserstein distances} \label{subsec:SWD}
For a thorough comparison of GANs, one needs to inspect a large number of fake samples. Doing this manually, would be very tedious and subjective. Instead, one should compare the results using meaningful, quantitative metrics.

For this, we follow \cite{Karras_2018_progressive} and calculate the sliced Wasserstein distances (SWD) introduced in Section \ref{subsec:evaluation_methods} between fake data samples and real data samples on various scales. These scales are introduced by building a 5-level Laplacian pyramid \cite{pyramid} with resolutions $16 \times 16$, $32 \times 32$, $64 \times 64$, $128 \times 128$, $256 \times 256$. Each pyramid level corresponds to a specific spatial resolution. For each level, we compute the SWD between the training dataset and a generated fake dataset of the same size. More precisely, the SWDs are calculated between datasets of random $7 \times 7$ patches of the pyramid samples. The patches are pre-processed by normalizing each channel (i.e. $x$ and $y$ displacement) to mean 0 and standard deviation 1. To reduce uncertainty, we average the SWDs of ten runs with randomly sampled fake data. Since there are less unique patches for low resolutions, we adapt the number of random patches depending on the pyramid level. For the five resolutions, $16 \times 16$, $32 \times 32$, $64 \times 64$, $128 \times 128$, and $256 \times 256$, we use 128, 256, 512, 1024, 2048 patches, respectively. The integral in Equation \eqref{eq:SWD} is approximated by choosing 512 random slices and averaging the results. We implemented a GPU-enabled version of the code from \cite{Karras_2018_progressive} using the PyTorch \cite{PyTorch} framework.

At least intuitively, a small SWD shows that the fake and real samples are similar. At low resolution (e.g. $16 \times 16$) only large-scale features like the crack length are visible and a small SWD would indicate that the variation of crack lengths in the fake dataset is similar to the training dataset. At high resolution (e.g. $256 \times 256$) very fine-grained structures like the inherent DIC noise is encoded in the patches. 

\begin{figure}
    \centering
    \includegraphics[width=0.85\textwidth]{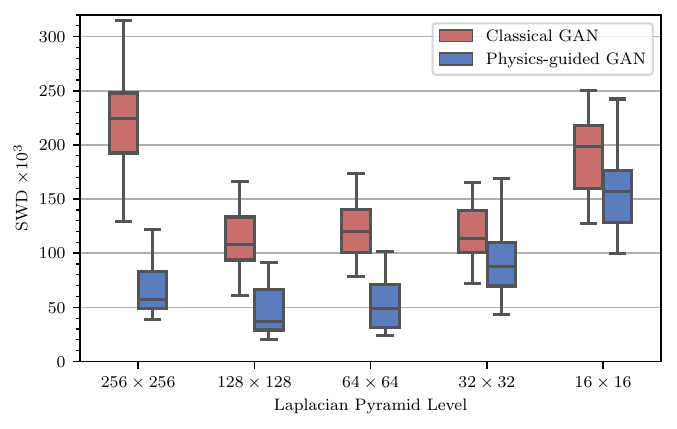}
    \caption{Comparison of SWDs between classical GAN (left) and physics-guided GAN (right) trained for 100 epochs. The boxplot intervals range from the minimal to the maximal SWDs. The box includes ranges from the 25\% to the 75\% quantile and shows the median.}
    \label{fig:swd_e100}
\end{figure}

Figure \ref{fig:swd_e100} shows the calculated SWDs of classical and physics-guided GAN samples after 100 epochs of training. In order to estimate uncertainty, we trained 10 randomly initialized models each with the classical and the physics-guided GAN architecture. The main observation is that for all resolutions, physics-guided samples are closer to the training data than classical GAN samples. This indicates that physics-guided GAN samples are better in quality and variation. Especially for the high resolution $256 \times 256$, the SWDs show a large gap and confirm our visual observation of artefacts and unphysical noise as seen in the classical GAN samples in Figure \ref{fig:samples_e100}. Nevertheless, the results can be significantly different for each trained generator. This fact is reflected in the large error bars of the SWDs.

\begin{figure}
    \centering
    \includegraphics[width=0.85\textwidth]{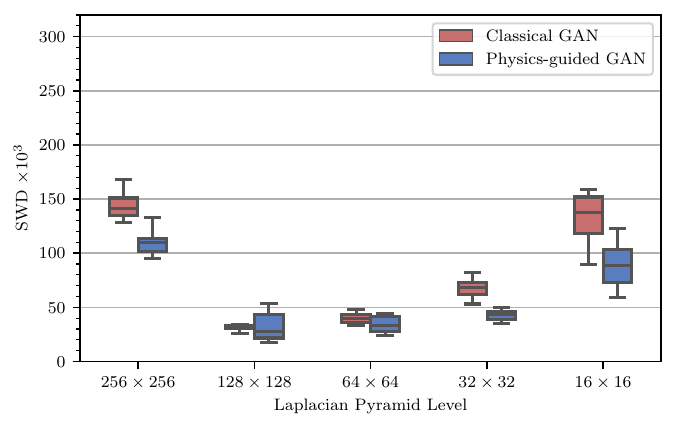}
    \caption{Comparison of SWDs between classical GAN (left) and physics-guided GAN (right) trained for 1000 epochs. The boxplot intervals range from the minimal to the maximal SWDs. The box includes ranges from the 25\% to the 75\% quantile and shows the median.}
    \label{fig:swd_e1000}
\end{figure}

The results after 1000 training epochs are displayed in Figure \ref{fig:swd_e1000}. Here, we used 2 training runs for each GAN architecture. As expected, the distances are all smaller than after 100 epochs. In contrast to the results after 100 epochs (cf. Figure \ref{fig:swd_e100}), both GAN architectures are closer together. However, the physics-guided GAN samples have significantly smaller SWDs for the fine resolution $256 \times 256$ and the low resolutions $16 \times 16$ and $32 \times 32$. This suggests that after 1000 epochs of training the physics-guided samples are still closer to the real samples in terms of quality and variation. Nevertheless, the few garbage samples seen in Figure \ref{fig:samples_e1000} \textcircled{2} could influence the SWDs especially at larger pyramid levels, e.g. $256 \times 256$.

\subsection{Geometry scores} \label{subsec:GS}
To compare the geometry score (GS) introduced in Section \ref{subsec:evaluation_methods} of different trained GANs, we generated fake datasets with the same number of samples $N=838$ as the training dataset. To calculate the MRLTs of the real and fake datasets, we mainly follow the recommendations in \cite{GeometryScore}. We set $i_{\rm max}=100$ and use $n=1000$ random landmarks. The number of samples in each landmark is $64$. The maximal persistance time $\alpha_{\rm max}$ is proportional to the maximal pairwise Euclidean distance between samples in each landmark, i.e. for $j = 1, \dots, n$:
\begin{equation}
    \alpha_{\rm max}^{j} = \gamma \max({\rm dist}(L_j,L_j)), \quad \gamma = \frac{1}{128}/\frac{N}{5000}.
\end{equation}
We used the implementation from \cite{GeometryScore} to calculate the MRLTs.

\begin{figure}
    \centering
    \includegraphics[width=0.85\textwidth]{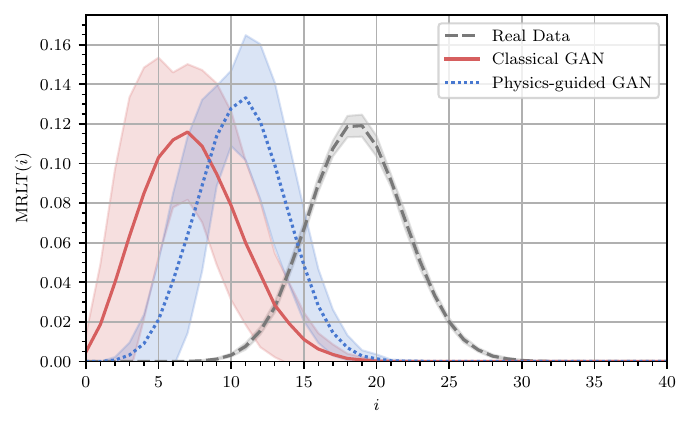}
    \caption{Comparison of MRLT distributions between the real dataset and fake datasets generated by classical and physics-guided GAN after 100 epochs of training.}
    \label{fig:mrlt_e100}
\end{figure}

Figure \ref{fig:mrlt_e100} shows the distributions of MRLTs after 100 epochs of training. The error band originates from the uncertainty induced by the random landmarks and, even more so, from the 10 different models trained for each GAN architecture. This results in large variations of calculated MRLTs. Nevertheless, on average the phyiscs-guided GAN distribution is closer to the MRLTs of the real data distribution than the classical GAN distribution. This observation is quantitatively reflected in a smaller mean GS of the phyiscs-guided models (see Table \ref{table:summary_results}). However, both fake data distributions are still far away from the real data distribution and the GSs are large.

\begin{figure}
    \centering
    \includegraphics[width=0.85\textwidth]{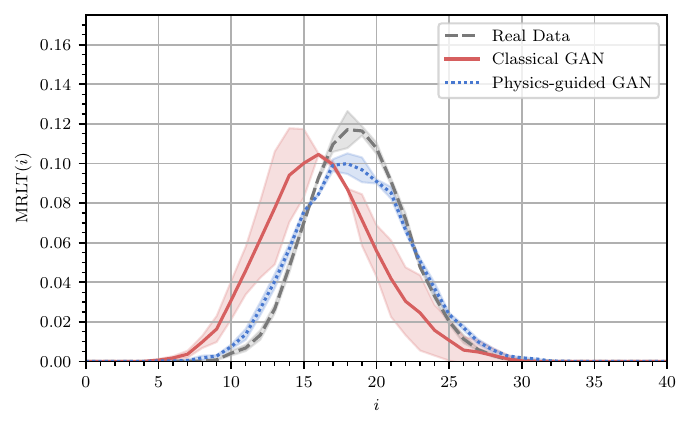}
    \caption{Comparison of MRLT distributions between the real dataset and fake datasets generated by classical and physics-guided GAN after 1000 epochs of training.}
    \label{fig:mrlt_e1000}
\end{figure}

In Figure \ref{fig:mrlt_e1000}, we see the MRLT distributions after 1000 epochs of training. Both GAN results are much closer to the real data than after 100 epochs and the phyiscs-guided GAN MRLTs almost coincide with the real data MRLTs. This accordance is shown in the calculated GSs in Table \ref{table:summary_results} as well.

\section{Conclusion} \label{sec:conclusion}
We introduced a machine learning framework to generate artificial full-field displacements of cracked specimen by learning the underlying data distribution from a sufficiently large digital image correlation dataset. The training data was obtained during fatigue crack growth experiments of the aluminium alloy AA2024-T3. 
In contrast to finite element simulations, our method is able to produce large amounts of interpolated DIC-like displacement data in a fast and easy way but is limited in the sense that boundary conditions and crack configurations cannot be controlled. 

Our approach is based on deep convolutional generative adversarial networks (DC-GANs). The main novelty compared to the classical DC-GAN framework is a physics-guided discriminator. This discriminator, in addition to the generated $x$- and $y$- displacement fields, gets also the derived \textsc{von Mises} equivalent strain as input. This enables the discriminator to detect physical inconsistencies in the generated fake samples more easily, thus enhancing the training process.

In order to evaluate trained generator models on an objective basis, we used two quantitative metrics. First, the sliced Wasserstein distance (SWD) between real and fake samples and, secondly, the geometry score (GS) approximating the topological distance between a generated data manifold and the training data manifold. 

We observed superior performance of the physics-guided GAN compared to the classical GAN approach. This result was observed by visual evaluation of generated samples and confirmed by lower SWDs and GSs of the physics-guided models. Both, SWD and GS, proved themselves to be valuable evaluation metrics. They are useful to identify mode collapse and to select the best trained models. Nevertheless, it is important to note that there is no natural metric to evaluate the performance of GANs. In the absence of powerful pre-trained models like Inception for DIC-like data, we had to stick to GAN metrics that are independent of these benchmark models. Our findings support the claim that hybrid models, which combine data-driven methods with physical domain knowledge, can lead to more powerful models and faster training.  

The visual inspection revealed a varying sample quality. Especially the converged models after 1000 epochs of training, apart from mostly good samples, produce few garbage samples. Although the number of these garbage samples is model-dependent, we were not able to avoid their occurrence completely. Moreover, we still face the issue of (local) non-convergence and mode collapse. To overcome these issues, one could try to stabilize training using suitable regularization techniques \cite{Mescheder2018ICML, Roth17}. 

The main open problem concerns the control of boundary conditions like the crack path and external force. In contrast to FE-based data generation, with our approach it is not possible to control them. This challenge could be tackled by using a conditional GAN framework \cite{cGAN2014} and is part of current research.

\section{Acknowledgements}
We acknowledge the financial support of the DLR-Directorate Aeronautics.

\section{Data availability}
The code and training data will be publicly available on Github (\url{https://github.com/dlr-wf}) and Zenodo (\url{https://doi.org/10.5281/zenodo.7737880}).

\section{Competing interests}
The Authors declare no Competing Financial or Non-Financial Interests.

\section{Author contributions}
D.M. and E.S. conceived the physics-guided GAN. E.S. implemented the neural network architectures, training algorithms, and evaluation metrics. All Authors discussed, analyzed, and interpreted the results and wrote the manuscript.

\printbibliography

@article{Tavares2017,
author = {Tavares, S. M. O. and de Castro, P. M. S. T.},
title = {An overview of fatigue in aircraft structures},
journal = {Fatigue \& Fracture of Engineering Materials \& Structures},
volume = {40},
number = {10},
pages = {1510-1529},
doi = {https://doi.org/10.1111/ffe.12631},
year = {2017}
}

@article{sutton2015recent,
  title={Recent advances and perspectives in digital image correlation},
  author={Sutton, M. A. and Hild, Fran{\c{c}}ois},
  journal={Experimental Mechanics},
  volume={55},
  number={1},
  pages={1--8},
  year={2015},
  publisher={Springer}
}

@article{Roux2009DIC,
  title={Digital image correlation and fracture: an advanced technique for estimating stress intensity factors of 2D and 3D cracks},
  author={Roux, St{\'e}phane and R{\'e}thor{\'e}, Julien and Hild, Fran{\c{c}}ois},
  journal={Journal of Physics D: Applied Physics},
  volume={42},
  number={21},
  pages={214004},
  year={2009},
  publisher={IOP Publishing}
}

@article{Becker2012DIC,
  title={An approach to calculate the J-integral by digital image correlation displacement field measurement},
  author={Becker, TH and Mostafavi, M and Tait, RB and Marrow, TJ},
  journal={Fatigue \& Fracture of Engineering Materials \& Structures},
  volume={35},
  number={10},
  pages={971--984},
  year={2012},
  publisher={Wiley Online Library}
}

@article{Zhao2019,
  title={The state of the art of two-dimensional digital image correlation computational method},
  author={Zhao, Jianlong and Sang, Yong and Duan, Fuhai},
  journal={Engineering reports},
  volume={1},
  number={2},
  pages={e12038},
  year={2019},
  publisher={Wiley Online Library}
}

@article{Melching2022,
  title={Explainable machine learning for precise fatigue crack tip detection},
  author={Melching, David and Strohmann, Tobias and Requena, Guillermo and Breitbarth, Eric},
  journal={Scientific Reports},
  volume={12},
  number={1},
  pages={9513},
  year={2022},
  publisher={Nature Publishing Group UK London}
}

@article{Karpatne17TGDS,
  title={Theory-guided data science: A new paradigm for scientific discovery from data},
  author={Karpatne, Anuj and Atluri, Gowtham and Faghmous, James H and Steinbach, Michael and Banerjee, Arindam and Ganguly, Auroop and Shekhar, Shashi and Samatova, Nagiza and Kumar, Vipin},
  journal={IEEE Transactions on knowledge and data engineering},
  volume={29},
  number={10},
  pages={2318--2331},
  year={2017},
  publisher={IEEE}
}

@misc{Daw17PGNN,
  doi = {10.48550/ARXIV.1710.11431},
  url = {https://arxiv.org/abs/1710.11431},
  author = {Daw, Arka and Karpatne, Anuj and Watkins, William and Read, Jordan and Kumar, Vipin},
  title = {Physics-guided Neural Networks (PGNN): An Application in Lake Temperature Modeling},
  publisher = {arXiv},
  year = {2017}
}

@article{Karniadakis2021PINNs,
  title={Physics-informed machine learning},
  author={Karniadakis, George Em and Kevrekidis, Ioannis G and Lu, Lu and Perdikaris, Paris and Wang, Sifan and Yang, Liu},
  journal={Nature Reviews Physics},
  volume={3},
  number={6},
  pages={422--440},
  year={2021},
  publisher={Nature Publishing Group UK London}
}

@article{Yang2019PIGGAN,
  title={Adversarial uncertainty quantification in physics-informed neural networks},
  author={Yang, Yibo and Perdikaris, Paris},
  journal={Journal of Computational Physics},
  volume={394},
  pages={136--152},
  year={2019},
  publisher={Elsevier}
}

@inproceedings{Daw21PIDGAN,
  title={PID-GAN: A GAN Framework based on a Physics-informed Discriminator for Uncertainty Quantification with Physics},
  author={Daw, Arka and Maruf, M and Karpatne, Anuj},
  booktitle={Proceedings of the 27th ACM SIGKDD Conference on Knowledge Discovery \& Data Mining},
  pages={237--247},
  year={2021}
}

@InProceedings{Erhan10UnsupervisedPretraining,
  title = 	 {Why Does Unsupervised Pre-training Help Deep Learning?},
  author = 	 {Erhan, Dumitru and Courville, Aaron and Bengio, Yoshua and Vincent, Pascal},
  booktitle = 	 {Proceedings of the Thirteenth International Conference on Artificial Intelligence and Statistics},
  pages = 	 {201--208},
  year = 	 {2010},
  editor = 	 {Teh, Yee Whye and Titterington, Mike},
  volume = 	 {9},
  series = 	 {Proceedings of Machine Learning Research},
  publisher =    {PMLR}
}

@InProceedings{Iscen2019CVPR,
author = {Iscen, Ahmet and Tolias, Giorgos and Avrithis, Yannis and Chum, Ondrej},
title = {Label Propagation for Deep Semi-Supervised Learning},
booktitle = {Proceedings of the IEEE/CVF Conference on Computer Vision and Pattern Recognition (CVPR)},
year = {2019}
}

@inproceedings{Goodfellow_GANs,
 author = {Goodfellow, Ian and Pouget-Abadie, Jean and Mirza, Mehdi and Xu, Bing and Warde-Farley, David and Ozair, Sherjil and Courville, Aaron and Bengio, Yoshua},
 booktitle = {Advances in Neural Information Processing Systems},
 editor = {Z. Ghahramani and M. Welling and C. Cortes and N. Lawrence and K.Q. Weinberger},
 pages = {},
 publisher = {Curran Associates, Inc.},
 title = {Generative Adversarial Nets},
 volume = {27},
 year = {2014}
}

@InProceedings{Gatys_NeuralStyleTransfer,
author={Gatys, Leon A. and Ecker, Alexander S. and Bethge, Matthias},
booktitle={2016 IEEE Conference on Computer Vision and Pattern Recognition (CVPR)}, 
title={Image Style Transfer Using Convolutional Neural Networks}, 
year={2016},
volume={},
number={},
pages={2414-2423},
doi={10.1109/CVPR.2016.265}
}

@InProceedings{Karras_2018_progressive,
title={Progressive Growing of {GAN}s for Improved Quality, Stability, and Variation},
author={Tero Karras and Timo Aila and Samuli Laine and Jaakko Lehtinen},
booktitle={International Conference on Learning Representations},
year={2018},
url={https://openreview.net/forum?id=Hk99zCeAb},
}

@InProceedings{Karras_2020_CVPR,
author = {Karras, Tero and Laine, Samuli and Aittala, Miika and Hellsten, Janne and Lehtinen, Jaakko and Aila, Timo},
title = {Analyzing and Improving the Image Quality of StyleGAN},
booktitle = {Proceedings of the IEEE/CVF Conference on Computer Vision and Pattern Recognition (CVPR)},
year = {2020}
}

@article{MOKHTARISHIRAZABAD201611,
title = {Evaluation of crack-tip fields from DIC data: A parametric study},
journal = {International Journal of Fatigue},
volume = {89},
pages = {11-19},
year = {2016},
note = {Special Issue: Crack Tip Fields 3},
doi = {https://doi.org/10.1016/j.ijfatigue.2016.03.006},
author = {M. Mokhtarishirazabad and P. Lopez-Crespo and B. Moreno and A. Lopez-Moreno and M. Zanganeh},
}

@article{Strohmann2021,
author = {Strohmann, Tobias and Starostin-Penner, Denis and Breitbarth, Eric and Requena, Guillermo},
title = {Automatic detection of fatigue crack paths using digital image correlation and convolutional neural networks},
journal = {Fatigue \& Fracture of Engineering Materials \& Structures},
volume = {44},
number = {5},
pages = {1336-1348},
doi = {https://doi.org/10.1111/ffe.13433},
year = {2021}
}

@misc{DCGAN2015,
  doi = {10.48550/ARXIV.1511.06434},
  url = {https://arxiv.org/abs/1511.06434},
  author = {Radford, Alec and Metz, Luke and Chintala, Soumith},
  title = {Unsupervised Representation Learning with Deep Convolutional Generative Adversarial Networks},
  publisher = {arXiv},
  year = {2015}
}

@article{odena2016deconvolution,
  author = {Odena, Augustus and Dumoulin, Vincent and Olah, Chris},
  title = {Deconvolution and Checkerboard Artifacts},
  journal = {Distill},
  year = {2016},
  doi = {10.23915/distill.00003}
}

@inproceedings{leakyrelu,
  title={Rectifier Nonlinearities Improve Neural Network Acoustic Models},
  author={Maas, Andrew L and Hannun, Awni Y and Ng, Andrew Y},
  year={2013},
  booktitle={Proceedings of the 30th International Conference on Machine Learning}
}

@inproceedings{IS,
 author = {Salimans, Tim and Goodfellow, Ian and Zaremba, Wojciech and Cheung, Vicki and Radford, Alec and Chen, Xi and Chen, Xi},
 booktitle = {Advances in Neural Information Processing Systems},
 editor = {D. Lee and M. Sugiyama and U. Luxburg and I. Guyon and R. Garnett},
 pages = {},
 publisher = {Curran Associates, Inc.},
 title = {Improved Techniques for Training GANs},
 volume = {29},
 year = {2016}
}

@inproceedings{FID,
 author = {Heusel, Martin and Ramsauer, Hubert and Unterthiner, Thomas and Nessler, Bernhard and Hochreiter, Sepp},
 booktitle = {Advances in Neural Information Processing Systems},
 editor = {I. Guyon and U. Von Luxburg and S. Bengio and H. Wallach and R. Fergus and S. Vishwanathan and R. Garnett},
 pages = {},
 publisher = {Curran Associates, Inc.},
 title = {GANs Trained by a Two Time-Scale Update Rule Converge to a Local Nash Equilibrium},
 volume = {30},
 year = {2017}
}

@INPROCEEDINGS{Inception,
  author={Szegedy, Christian and Wei Liu and Yangqing Jia and Sermanet, Pierre and Reed, Scott and Anguelov, Dragomir and Erhan, Dumitru and Vanhoucke, Vincent and Rabinovich, Andrew},
  booktitle={2015 IEEE Conference on Computer Vision and Pattern Recognition (CVPR)}, 
  title={Going deeper with convolutions}, 
  year={2015},
  volume={},
  number={},
  pages={1-9},
  doi={10.1109/CVPR.2015.7298594}
}

@article{ImageNet,
  author = {Russakovsky, O. and Deng, J. and Su et al., H.},
  title = {ImageNet Large Scale Visual Recognition Challenge},
  journal = {Int J Comput Vis},
  year = {2015},
  volume = {115},
  pages = {211-252},
  doi = {10.1007/s11263-015-0816-y}
}

@misc{AdamArXiv,
  author = {Kingma, Diederik P. and Ba, Jimmy},
  title = {Adam: A Method for Stochastic Optimization},
  publisher = {arXiv},
  year = {2014},
  doi = {10.48550/arXiv.1412.6980}
}

@InProceedings{WGANs,
  title = 	 {{W}asserstein Generative Adversarial Networks},
  author =       {Martin Arjovsky and Soumith Chintala and L{\'e}on Bottou},
  booktitle = 	 {Proceedings of the 34th International Conference on Machine Learning},
  pages = 	 {214--223},
  year = 	 {2017},
  editor = 	 {Precup, Doina and Teh, Yee Whye},
  volume = 	 {70},
  series = 	 {Proceedings of Machine Learning Research},
  publisher =    {PMLR}
}

@misc{Arjovsky17,
  doi = {10.48550/ARXIV.1701.04862},
  url = {https://arxiv.org/abs/1701.04862},
  author = {Arjovsky, Martin and Bottou, Léon},
  title = {Towards Principled Methods for Training Generative Adversarial Networks},
  publisher = {arXiv},
  year = {2017}
}

@InProceedings{RabinWasserstein2012,
author="Rabin, Julien
and Peyr{\'e}, Gabriel
and Delon, Julie
and Bernot, Marc",
editor="Bruckstein, Alfred M.
and ter Haar Romeny, Bart M.
and Bronstein, Alexander M.
and Bronstein, Michael M.",
title="Wasserstein Barycenter and Its Application to Texture Mixing",
booktitle="Scale Space and Variational Methods in Computer Vision",
year="2012",
publisher="Springer Berlin Heidelberg"
}

@inbook{pyramid,
author = {Burt, Peter J. and Adelson, Edward H.},
title = {The Laplacian Pyramid as a Compact Image Code},
year = {1987},
publisher = {Morgan Kaufmann Publishers Inc.},
address = {San Francisco, CA, USA},
booktitle = {Readings in Computer Vision: Issues, Problems, Principles, and Paradigms},
pages = {671–679}
}

@InProceedings{GeometryScore,
  title = 	 {Geometry Score: A Method For Comparing Generative Adversarial Networks},
  author =       {Khrulkov, Valentin and Oseledets, Ivan},
  booktitle = 	 {Proceedings of the 35th International Conference on Machine Learning},
  pages = 	 {2621--2629},
  year = 	 {2018},
  editor = 	 {Dy, Jennifer and Krause, Andreas},
  volume = 	 {80},
  series = 	 {Proceedings of Machine Learning Research},
  publisher =    {PMLR}
}

@book{Goodfellow-et-al-2016,
    title={Deep Learning},
    author={Ian Goodfellow and Yoshua Bengio and Aaron Courville},
    publisher={MIT Press},
    year={2016}
}

@book{HatcherAlgebraicTopology,
      author        = "Hatcher, Allen",
      title         = "{Algebraic topology}",
      publisher     = "Cambridge Univ. Press",
      address       = "Cambridge",
      year          = "2000"
}

@inproceedings{Mescheder2018ICML,
  author = {Lars Mescheder and Sebastian Nowozin and Andreas Geiger},
  title = {Which Training Methods for GANs do actually Converge?},
  booktitle = {International Conference on Machine Learning (ICML)},
  year = {2018}
}

@inproceedings{Roth17,
author = {Roth, Kevin and Lucchi, Aurelien and Nowozin, Sebastian and Hofmann, Thomas},
title = {Stabilizing Training of Generative Adversarial Networks through Regularization},
year = {2017},
publisher = {Curran Associates Inc.},
address = {Red Hook, NY, USA},
booktitle = {Proceedings of the 31st International Conference on Neural Information Processing Systems},
pages = {2015–2025},
location = {Long Beach, California, USA},
series = {NIPS'17}
}

@misc{cGAN2014,
  doi = {10.48550/ARXIV.1411.1784},
  url = {https://arxiv.org/abs/1411.1784},
  author = {Mirza, Mehdi and Osindero, Simon},
  title = {Conditional Generative Adversarial Nets},
  publisher = {arXiv},
  year = {2014}
}

@inproceedings{ReLUBoltzmannMachines_paper,
	title={Rectified linear units improve restricted boltzmann machines},
	author={Nair, Vinod and Hinton, Geoffrey E},
	booktitle={Icml},
	year={2010}
}

@inproceedings{BatchNorm_paper,
	title={Batch normalization: Accelerating deep network training by reducing internal covariate shift},
	author={Ioffe, Sergey and Szegedy, Christian},
	booktitle={International conference on machine learning},
	year={2015},
}

@incollection{PyTorch,
title = {PyTorch: An Imperative Style, High-Performance Deep Learning Library},
author = {Paszke, Adam and Gross, Sam and Massa, Francisco and Lerer, Adam and Bradbury, James and Chanan, Gregory and Killeen, Trevor and Lin, Zeming and Gimelshein, Natalia and Antiga, Luca and Desmaison, Alban and Kopf, Andreas and Yang, Edward and DeVito, Zachary and Raison, Martin and Tejani, Alykhan and Chilamkurthy, Sasank and Steiner, Benoit and Fang, Lu and Bai, Junjie and Chintala, Soumith},
booktitle = {Advances in Neural Information Processing Systems 32},
pages = {8024--8035},
year = {2019},
publisher = {Curran Associates, Inc.}
}

@article{vonMises1913,
  title={Mechanik der festen K{\"o}rper im plastisch-deformablen Zustand},
  author={Mises, R v},
  journal={Nachrichten von der Gesellschaft der Wissenschaften zu G{\"o}ttingen, Mathematisch-Physikalische Klasse},
  volume={1913},
  pages={582--592},
  year={1913}
}

@article{Lava2009,
author = {P. Lava and S. Cooreman and S. Coppieters and M. {De Strycker} and D. Debruyne},
title = {Assessment of measuring errors in DIC using deformation fields generated by plastic FEA},
journal = {Optics and Lasers in Engineering},
volume = {47},
number = {7},
pages = {747-753},
year = {2009}
}

@article{Rohe2022,
author = {Rohe, D. and Jones, E.},
year = {2022},
pages = {615-631},
title = {Generation of Synthetic Digital Image Correlation Images Using the Open-Source Blender Software},
volume = {46},
journal = {Experimental Techniques}
}

@article{Balcaen2017,
author = {Balcaen, R. and Wittevrongel, L. and Reu, P. L. and Lava, P. and Debruyne, D.},
year = {2017},
pages = {703-718},
title = {Stereo-DIC Calibration and Speckle Image Generator Based on FE Formulations},
volume = {57},
journal = {Experimental Techniques}
}

@article{su_recursive-iterative_2020,
	title = {Recursive-iterative digital image correlation based on salient features},
	volume = {59},
	number = {3},
	journal = {Optical Engineering},
	author = {Su, Zhilong and Lu, Lei and He, Xiaoyuan and Yang, Fujun and Zhang, Dongsheng},
	year = {2020},
	pages = {034111}
}

@article{su_refractive_2021,
	title = {Refractive three-dimensional reconstruction for underwater stereo digital image correlation},
	volume = {29},
	number = {8},
	journal = {Optics Express},
	author = {Su, Zhilong and Pan, Jiyu and Lu, Lei and Dai, Meiling and He, Xiaoyuan and Zhang, Dongsheng},
	year = {2021},
	pages = {12131}
}

@article{Chen2018,
	title = {Optimized digital speckle patterns for digital image correlation by consideration of both accuracy and efficiency},
	volume = {57},
	number = {4},
	journal = {Applied Optics},
	author = {Chen, Zhenning and Shao, Xinxing and Xu, Xiangyang and He, Xiaoyuan},
	year = {2018},
	pages = {884}
}

\end{document}